\documentclass[letterpaper]{article} 
\usepackage{aaai23}  
\usepackage{times}  
\usepackage{helvet}  
\usepackage{courier}  
\usepackage[hyphens]{url}  
\usepackage{graphicx} 
\usepackage{natbib}  
\usepackage{caption} 
\usepackage{algorithm}
\usepackage{algorithmic}

\usepackage{newfloat}
\usepackage{listings}
\DeclareCaptionStyle{ruled}{labelfont=normalfont,labelsep=colon,strut=off} 
\floatstyle{ruled}
\newfloat{listing}{tb}{lst}{}
\floatname{listing}{Listing}
\usepackage[table]{xcolor}

\begin{document}
%
\title{Hybrid Navigation Acceptability and Safety}
\author {
    Benoit Clement\textsuperscript{\rm {1,2,3}},
    Marie Dubromel\textsuperscript{\rm 1},
    Paulo E. Santos\textsuperscript{\rm {1,3}},
    Karl Sammut\textsuperscript{\rm {1,3}},
    Michelle Oppert\textsuperscript{\rm 1,\rm 4},
    Feras Dayoub\textsuperscript{\rm 1,\rm 5}
}
\affiliations {
    \textsuperscript{\rm 1} CNRS IRL 2010 CROSSING, Adelaide, South Australia\\
    \textsuperscript{\rm 2} ENSTA Bretagne, Brest, France\\
    \textsuperscript{\rm 3} Flinders University, Adelaide, South Australia\\
    \textsuperscript{\rm 4} University of South Australia, Adelaide, South Australia\\
    \textsuperscript{\rm 5} University of Adelaide, Adelaide, South Australia\\
    benoit.clement@ensta-bretagne.fr
}

\maketitle
\begin{abstract}
\begin{quote}
Autonomous vessels have emerged as a prominent and accepted solution, particularly in the naval defence sector. However, achieving full autonomy for marine vessels demands the development of robust and reliable control and guidance systems that can handle various encounters with manned and unmanned vessels while operating effectively under diverse weather and sea conditions. A significant challenge in this pursuit is ensuring the autonomous vessels' compliance with the International Regulations for Preventing Collisions at Sea (COLREGs). These regulations present a formidable hurdle for the human-level understanding by autonomous systems as they were originally designed from common navigation practices created since the mid-19th century. Their ambiguous language assumes experienced sailors' interpretation and execution, and therefore demands a high-level (cognitive) understanding of language and agent intentions. These capabilities surpass the current state-of-the-art in intelligent systems. This position paper highlights the critical requirements for a trustworthy control and guidance system, exploring the complexity of adapting COLREGs for safe vessel-on-vessel encounters considering autonomous maritime technology competing and/or cooperating with manned vessels.
\end{quote}
\end{abstract}

\section{Introduction}
\noindent Autonomous vessels are rapidly gaining acceptance, particularly within the naval defence sector, as they offer an obvious means of removing human personnel from risks originating from conflict or environmental threats. A critical requirement of achieving full autonomy for a marine vessel is the development of a robust and trustworthy control and guidance system that accommodates different approach encounters (considering manned as well as unmanned vessels), while operating under a wide range of weather and sea state conditions. To safely accommodate vessel-on-vessel encounters, the autonomous vessel must comply with the International Regulations for Preventing Collisions at Sea (COLREGs) \cite{imo1972convention,2012CL}.  COLREGs evolved from a set of practises that were originally designed in the mid-19th century for human interpretation. Therefore, these rules are written in ambiguous prose, assuming that their interpretation and execution were carried out by highly experienced sailors and not by an autonomous system. Here is a list of the relevant COLREGs rules that are considered in our work:
\begin{itemize}
    \item Rule 8 - \textit{Actions to avoid collision}: if there is sufficient sea-room, alteration of course alone may be most effective. Reduce speed, stop or reverse only if necessary. 
    \item Rule 13 - \textit{Overtaking}: Any vessel overtaking any other shall keep out of the way of the vessel being overtaken.
    \item Rule 14 - \textit{Head-on}: Each head-on vessel shall alter her course to starboard so that each shall pass on the port side of the other.
    \item Rule 15 - \textit{Crossing}: The vessel which has the other on her own starboard side shall keep out of the way.
    \item Rule 16 - \textit{Actions by give-way vessel}: Take early and substantial action to keep well clear. 
    \item Rule 17 - \textit{Actions by stand-on vessel}: Keep her course and speed but may take action to avoid collision if the other vessel is not taking a COLREGs-compliant action.
\end{itemize}

Moreover, while implementing COLREGs in an environment exclusively inhabited by unmanned vessels is relatively straightforward, its implementation becomes significantly more complex in an environment where unmanned vessels interact with manned ones. This complexity arises from the often unpredictable behavior of human navigators, who may occasionally deviate from the rules in their efforts to avoid potentially hazardous situations. In contrast, current COLREG-compliant methods strictly adhere to the rules, regardless of the navigator's intentions. Consequently, in a potential future scenario, fleets of autonomous vessels could be diverted from their intended course or even hijacked by malicious actors manipulating these machines' nearly explicit knowledge states. Developing systems capable of such advanced epistemic reasoning, particularly in mixed-motive situations, is one of the primary objectives of this research. 

Classical model-based approaches to automated COLREG compliance have proven to be too complicated to accommodate all possible encounters, environment scenarios, and human behaviours \cite{statheros_howells_maier_2008}. Modern Machine Learning (ML) methods, such as Deep Reinforcement Learning (DRL), could provide a flexible and adaptable model-free guidance and obstacle avoidance system, whereby the multiple possible interactions and scenarios can be abstracted from previous observations \cite{burmeister21}. However, ML methods do not provide the semantics of rules, or possible ways of breaking them, given the situation perceived. This work proposes the development of an AI-based  COLREG-compliant model-free collision avoidance system that will be trained using historical AIS-based simulations of real-world scenarios, whereas the possible human interpretations of the rules will take a centre point in the development. This will be accomplished according to the following four sub-modules:

\begin{itemize}
    \item Module 1: Autonomous Situation Awareness (ASA) aims to classify obstacles, other vessels, and intentions, defining vessel-on-vessel encounters from the information provided by multiple sensors (e.g. AIS and radar) .
    \item Module 2: Readability of Human Rules (RHR) will take into account how to create algorithms capable of presenting a {\em human-like} understanding of the COLREG rules, which should take into account the multiple possible space-time histories consistent with the observations and inferences provided by the ASA subsystem.
    \item Module 3: Path Planning and Control (PPC) aims at the implementation of guidance and control algorithms to ensure acceptability and safety based on human understanding of the COLREGs provided by the RHR subsystem.
    \item Module 4: Human Acceptability (HA) is the task of measuring the acceptability of the behaviour of autonomous systems based on studies of human operators.
\end{itemize}

\section{Module 1: Autonomous Situation Awareness}
The Autonomous Situation Awareness (ASA) module in this work has the task of integrating data from multiple sensors, including radar, Automatic Identification System (AIS) and possibly cameras and Synthetic Aperture Radar (SAR) imagery. Using machine learning techniques, the ASA system's task is to accurately identify and track vessels, obstacles, and navigational hazards to provide a comprehensive interpretation of the situation in which the vessel is emerged.

Developing an ASA system is crucial to ensure the safe operation of Autonomous Surface Vessels (ASV) in open-water environments. By providing real-time situational awareness of the vessel's surroundings and predicting potential risks, the ASA system will significantly enhance the safety and reliability of autonomous navigation. This information will be used by the PPC system (module 3) to generate safe and COLREG-compliant trajectories and vessel behaviour.

The proposed ASA system should have the following features:
\begin{itemize}
    \item Object detection and tracking:  using a combination of sensors, a system will be implemented to detect and track maritime objects, including other vessels, buoys, and obstacles. This information will be used to adjust the vessel's course and speed to avoid collisions.
    \item Real-time situational awareness: the proposed ASA system will process data from various sensors and provide real-time situational awareness to the vessel control system. This should allow the vessel to make informed decisions and take appropriate actions in a dynamic and changing environment.
    \item Adaptive decision-making: the proposed ASA system will use machine learning algorithms to fuse sensor data and make adaptive decisions based on the vessel's goals, objectives, and COLREGs.
\end{itemize}

The main research questions that should be considered during this development are the following: 

\begin{enumerate}
    \item What are the requirements for an ASA system for autonomous navigation of ASVs in busy open waters?
    \item How can machine learning techniques be used to accurately identify and track other vessels, obstacles, and navigational hazards in real-time?
    \item How can the ASA system be integrated with the ASV's collision avoidance planner to generate safe and compliant trajectories that comply with the Convention on the International Regulations for Preventing Collisions at Sea (COLREGs)? 
    \item How can the performance of the ASA system be tested and validated in various scenarios, including heavy traffic and adverse weather conditions?
\end{enumerate}

\section{Module 2: Readability of Human Rules}
Laws and regulations, such as COLREGs, are inherently rule-based.  They invariably state constraints that must be followed or activities for which permission or obligations are given.  These rule-based conventions that govern the behaviour of entities in the world need to be captured so that robotic and autonomous systems do not violate them. While machine learning (ML) approaches can likely capture some intended constraints on behaviour, given enough effort on creating training examples, the uncertainly in the quality of a ML output does not support an acceptable level of trust in any ML-based system. For instance, having an autonomous vehicle that only takes action to avoid a crash 90\% of the time, while acceptable for a machine learning academic work, it is unacceptable in real situations.  In many cases, it may also be impractical or at least inefficient to generate a statistically significant set of training examples for every possible relevant scenario.  In this context, we propose a combination of ML algorithms with approaches for representing reasoning about COLREGS, such as employing the Suggested Upper Merged Ontology (SUMO) \cite{sumo01}, which embodies two decades of work on a reusable inventory of common concepts.

In general, existing collision avoidance algorithms translate COLREGS as a set of hard navigational constraints (e.g. \cite{statheros_howells_maier_2008}), ignoring the navigator’s intentions and potential misunderstanding (or misuse) of the rules intended semantics. Past research has been devoted to the formal representation of a limited set of COLREG rules in terms of ontologies \cite{kreutzmann2013towards,Dylla09}. However, this previous research has used a very limited notion of high-level formalisation (akin to a taxonomy rather than to a well-defined, formalised, body of knowledge), and the rules defined were written independently from the formalisation of other important concepts (such as those related to meteorology, or sea conditions). Moreover, little attention has been given to the human understanding of these rules and, to the best of our knowledge, there was no mention in the current scientific literature about how to interpret a ship’s behaviour, considering the navigator's intentions. Therefore, the present proposal aims at bridging the gap between the human interpretation of COLREG rules, situation awareness (provided by ML algorithms) and real-time path planning systems. As ever larger commercial ships are provided leaner crews, the necessity of ship automation becomes more pressing.  This proposal has the potential to accelerate the transfer of autonomy from at least some human-guided vehicles to machine-guided scenarios.

\section{Module 3: Path Planning and Control}

This part of the project aims to develop and explore multi-constraint optimisation-based planners \cite{lan2020,TSOLAKIS2022269} that can efficiently identify long-term trajectories for Autonomous Surface Vehicles (ASV) while ensuring compliance with the human-understanding of COLREGs. To achieve these objectives, the planners will utilise multi-constraint optimisation techniques. This approach involves finding the best trajectory for the ASV, considering multiple constraints that may be related to the vehicle's dynamics, environmental conditions, and most importantly, adherence to the COLREGs formalisation obtained in Module 2.
In order to anticipate future collisions and plan safe trajectories, the planners will conduct joint forward simulations of both the ASV and other manned/unmanned vessels operating in the region. By simulating the movements of all vessels together, potential collision scenarios can be identified and avoided.
Virtual obstacles will be constructed to represent the constraints imposed by COLREGs during the optimisation process. These obstacles will effectively encode the navigational rules and regulations specified in the COLREGs, ensuring that the trajectories generated by the planner adhere to international maritime regulations.
The evaluation of the proposed planners will be carried out in two parts. First, planners will be tested in single-ship encounters to demonstrate their ability to produce COLREG-compliant trajectories when encountering a single other vessel. Second, the planners will be compared against state-of-the-art methods in more complex scenarios involving multi-ship encounters, whereas the simulation will use real descriptions of ship-on-ship encounters (including edge cases in which distinct interpretations of the rules could be applied). These multi-ship encounters pose greater challenges, as the planners must navigate through potentially crowded and dynamic environments, with autonomous, semi-autonomous and manual ships, while avoiding collisions and strictly adhering to the COLREGs.

The success rates of the proposed planners will be defined as key performance metrics. These metrics should measure the planners' efficiency and effectiveness in generating trajectories that comply with the COLREGs, according to the general human understanding of these rules. The success rates may include the percentage of successful COLREGs-compliant trajectories generated in different scenarios, the average time taken to find a feasible trajectory, or the overall safety and collision avoidance performance in both single and multi-ship encounters.

To facilitate proactive collision avoidance, autonomous vessels will need to be able to make long-term trajectory predictions, taking into account the situation awareness inferences output by the algorithms developed in Module 1, and also by artificially representing the navigation experience while emulating the human mental models that facilitate these functions. We suggest taking advantage of Machine Learning to emulate the development of mental models that are constrained by (and consistent with respect to) the COLREG formalisations provided in Module 2. 

 
We list here the steps that can feed our methodologies to go from data to models.
\begin{enumerate}
    \item \textbf{AIS Data}.
    We propose to exploit historical Automatic Identification System (AIS) data to serve as a synthetic form of navigation experience. AIS relays information on ship behaviour, such as position, heading, speed, and ship type. By examining AIS histories, we can get an idea of the historical behaviour of the vessel. This can be seen as analogous to a navigator's experience of historical ship behaviour for a given geographical region. For the European zone, the AIS data recovered in 2016 give the map shown in Figure \ref{fig:AIS}. Current work at the Ecole Navale \cite{ELAYAM22} can be used as a basis for the study. 
To facilitate long-term trajectory predictions, it is desirable to develop methods to emulate the development of mental models. Thus, methods must first be developed to classify data into categories of specific ship behaviour. Next, methods to facilitate the matching of trajectory segments to a category of ship behaviour should be investigated. Finally, methods to model the dynamics within each category of ship behaviour need to be developed. In this way, a new trajectory can be assigned to a given category, and its future trajectory can be predicted based on the unique behaviour of that category.

\begin{figure}
    \centering
\includegraphics[width=\linewidth]{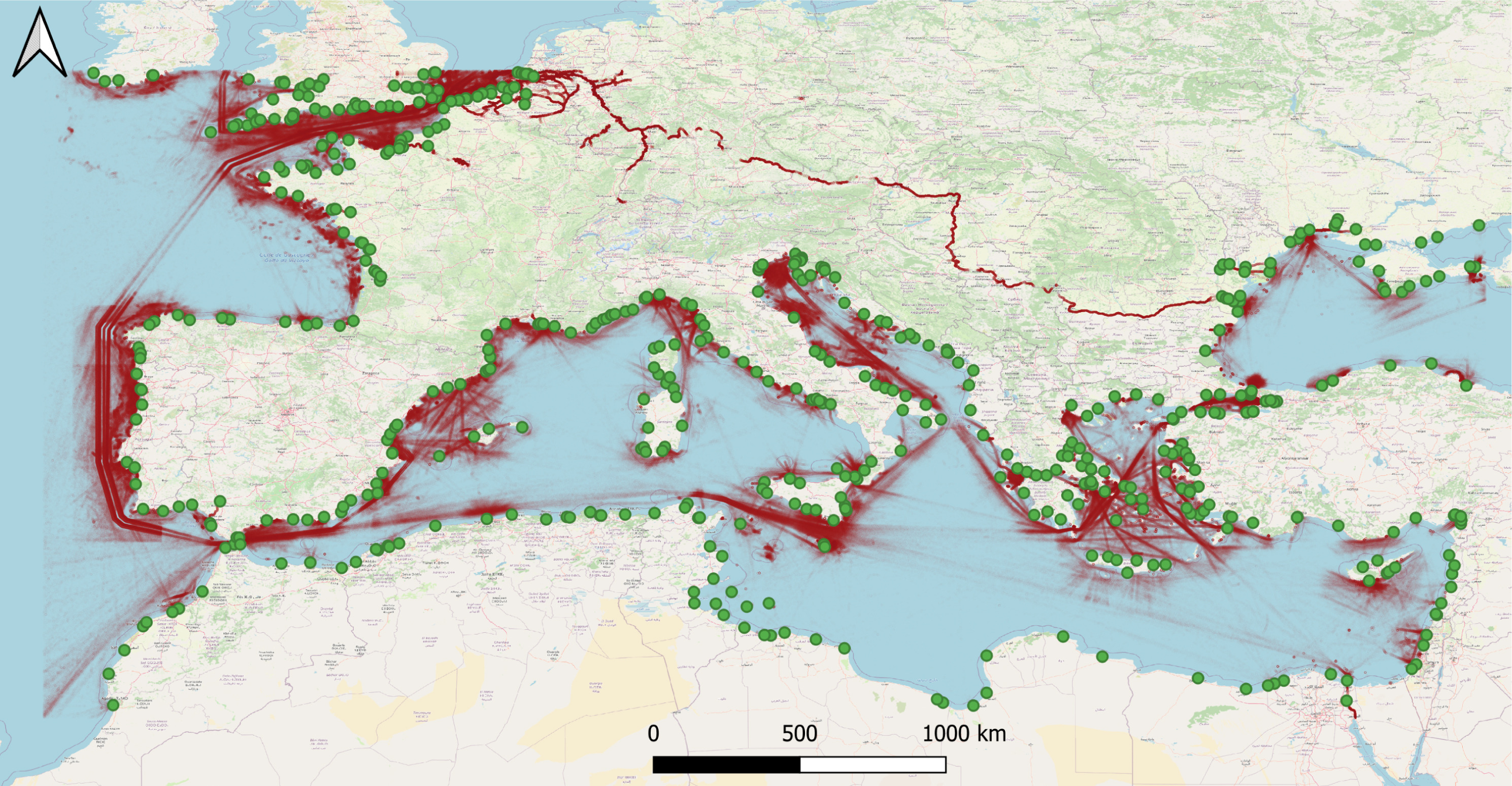} 
\caption{Illustration of maritime traffic based on AIS data in Europe (from \cite{ELAYAM22}).}
    \label{fig:AIS}
\end{figure}

The development of this part of the project should follow the sequence of steps listed below:

\item \textbf{Clustering}. The first step in mimicking the development of mental models is to facilitate categorisation functions. To predict future ship trajectories, historical ship behaviour needs to be decomposed into groups of specific behaviours. This is analogous to \textit{clustering}, where similar data are grouped into clusters. 

\item \textbf{Classification} The second step is to facilitate the mapping of a new trajectory to one of the existing categories of ship behaviour. Model matching will facilitate the selection of the appropriate category used to understand the situation, as well as the use of the corresponding behaviour model to project future dynamics. This is therefore classification in the sense of Machine Learning.

\item \textbf{Prediction}. To facilitate the prediction of long-term trajectories, each cluster must have a behavioural model. This mirrors the same functionality of mental models, where each category has a transition model capable of predicting future dynamics. Machine Learning is also capable of facilitating such regression functions. 

\item \textbf{Deep Learning to emulate high-level \textit{situation awareness}.}
In the longer term, Deep Learning and AIS for ship trajectory prediction seem to be a promising development path. It could be proposed that historical AIS data for a given region be aggregated to reflect specific historical behaviour. 


\end{enumerate}

In this study, we will attempt to show that proactive collision avoidance actions taken some time before the nearest point of approach can make the cohabitation of autonomous and non-autonomous vessels on a given stretch of water safer and more fluid. The aim of taking into account the behaviours recorded by the AIS is to allow conventional vessels and autonomous vehicles to co-habit.
We will seek to develop models through a classification of ship behaviour, where each class will have a specific transition function that models future dynamics. When the models are matched, a new trajectory will be able to fall into one of the existing classes, and the appropriate model will be applied to predict the future behaviour of a given vessel.


\section{Module 4: Human Acceptability}
This module aims to investigate human factors associated with the human acceptance of autonomous maritime vessels in following a set of rules of navigation / interaction (COLREGS in the case of a maritime application). These salient human factors include situational awareness, decision-making, workload changes, skills and training requirements, trust and reliance, and ethical considerations. Understanding stakeholders (currently humans who adhere to COLREGs) and their acceptance of autonomous systems of their vessels is of interest for future research and will inform the development of guidelines and integration of advanced autonomous systems and agents in maritime environments.

In the context of humans and AI, adoption and adaption are related but distinct concepts.
Adoption refers to the decision to implement or use AI technologies. In other words, if the human takes on AI. This involves deciding to implement AI technologies in processes and could include purchasing of developing AI systems. Adoption is considered a strategic decision that considers factors such as cost, benefits, and risk, but can also consider perceptions by potential users. 
Adaption refers to the practical actions and strategies taken to integrate and interact with these technologies effectively. In other words, how the human uses AI. This involves practical steps to integrate and interact with AI systems, including developing data management protocols, training other users, and monitoring the performance of the system. Adaption is an ongoing process that requires continuous evaluation and adjustment to ensure that AI systems remain effective and aligned with the users' objectives. 
Acceptance underpins the adoption and adaption of AI technologies. Without acceptance, users may resist or be hesitant to use AI technologies, which can limit their effectiveness and potential benefits. It could be considered the foundation for any effort to implement and integrate AI technologies effectively. 
Promoting the acceptance of AI technologies can also help address any concerns or fears users may have about AI’s impact on employment, privacy, and society (for example). Building trust and understanding of AI technologies means that users can better leverage the benefits of these technologies while also mitigating any potential risks or concerns. 
Understanding acceptance may be iterative depending on adoption and adaption and needs to be further understood. 


The use of autonomous systems in the context of COLREGs introduces new challenges related to legal and regulatory frameworks, liability and responsibility, cybersecurity, and human factors. For example:
\begin{itemize}
    \item clear guidelines and regulations to govern the use, and consequences, of COLREGs  in the context of autonomous systems;
    \item considerations of liability and responsibility in the event of an incident or accident. Who would be held responsible – particularly if there was zero crew? 
    \item Autonomous systems rely on software and communication technologies, so, this can make them vulnerable to cyber-attacks. Cybersecurity is an increasingly important issue in all realms, not least maritime;
    \item An autonomous system could help improve safety and efficiency, but they can also introduce new risks related to human factors such as situational awareness or monitoring of systems and performance.
\end{itemize}
Human factors are important when considering Adoption, Adaption, and Acceptance (AAA). Some of these reasons include perceived challenges or changes in situational awareness, familiarity (or unfamiliarity) in the operational context, changes in workload, skills and training requirements, trust and reliance, and ethical considerations. Autonomous systems could have a meaningful impact on human factors in maritime operations, and it is important to address these factors in design and implementation of autonomous systems to ensure safe and effective integration into maritime operations, in other words, to increase adoption, adaption, and overall acceptance:
\begin{itemize}
    \item Adopting autonomous systems aboard maritime vessels may be interpreted by users in opposing ways. Initial research should aim to understand barriers and drivers for the perceived implementation of these systems. This process should feedback on design and development;
    \item Adaption should be tested and observed with users and considered with a feedback system between users and designers/developers;
    \item Acceptance should be understood to inform on policy change and other associated regulatory policies and procedures. 
\end{itemize}

These assessments are recommended to include a blend of methodologies to collect the most salient data, as these systems will directly impact stakeholders. It is recommended that qualitative data such as interviews and focus groups are conducted throughout the AAA process. Quantitative data can be collected simultaneously to ensure congruence between what is said and observed, and what is thought. Validated instruments such as the Technology Acceptance Model (TAM) \cite{Davis89,Sohn20} could be adapted to autonomous systems throughout the AAA process to map changes in user perceptions. Paired with this, users are encouraged to participate in experiments from initial simulations to real-world observations across the AAA process. These experiments could include replications of real-world scenarios where decision-making, situational awareness, and workload changes are experienced with the use of the autonomous system (such as realistic scenarios, for instance, fatigue of crew due to illness, extreme weather, or high-stakes service-related demands). This would be useful for developers and stakeholders across the AAA process to encourage the final acceptance of autonomous systems. This process would highlight ethical issues and provide insights into the future skills and training required to continue to sail vessels but in combination with autonomous systems.


\section{First Simulations}

In order to measure the success rate of the development of an avoiding collision system for autonomous, semi-autonomous, and manual ships, a new light simulator has been developed. The main objective of this simulator is to both replay a past scenario based on collected AIS data, and add multiple ASV in the scene without causing any disturbance. This simulator also takes into account the rules of the COLREGs to avoid any kind of collision.

\begin{figure}
    \centering
\includegraphics[width=0.49\linewidth]{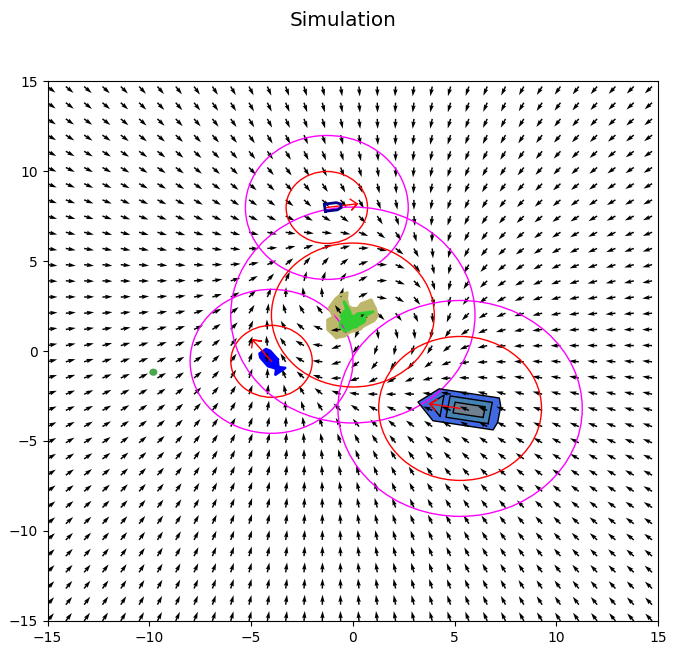} 
\includegraphics[width=0.5\linewidth]{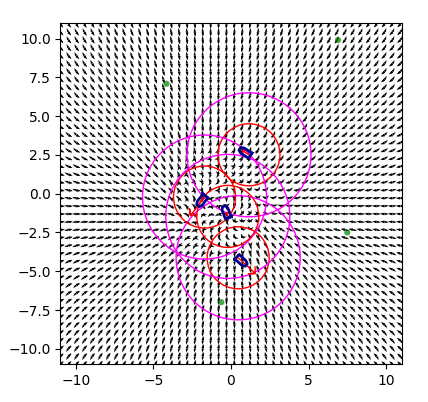} 
\caption{Illustrations of different simulation runned with different types of agents.}
\label{fig.agents}
\end{figure}

To include an ASV in maritime traffic without any disturbance, any ASV coming from the AIS data are considered as obstacles. Therefore, all added ASV must correct their initial trajectory to avoid collision with the AIS data AVS. To solve this issue, a privilege scale is introduced (from 0 to 1000). Before running the simulation, multiple ASV can be initialised, with a position ($x$ and $y$), a speed ($v$), a heading ($\theta$), but also with the type of ASV it is needed to create. Here are the different types of ASV that have been implemented as illustrated in Figure \ref{fig.agents}:

\begin{itemize}
    \item \textbf{Boat}: Representing a pleasure boat with a hull length less than or equal to 24 metres. Its privilege degree is set to \textbf{0} because it is the most mobile agent.
    \item \textbf{Ship}: Representing an ocean liner (couple of hundreds metres), to simulate bigger agents, slower with less manoeuvrability. Therefore, its privilege degree is set to \textbf{30}.
    \item \textbf{Whale}: Representing all the agents for which no information are known, so with a complete uncertainty about their future trajectory. It can represent marine animal, jet-skis, or any another agent too small to be registered or detected. Due to its unknown behaviour, its privilege degree will be set to \textbf{500}.
    \item \textbf{Island}: Representing any sort of non-mobile agent for which it is impossible to move and avoid a collision. It could be a natural obstacle part of the environment like an island, a rock, a reef, or an anchored agent. Therefore, its privilege degree is set to the maximum value \textbf{1000}.
\end{itemize}
The structure of the simulator offers the option of adding new types of maritime objects.

The ASV which will have to operate the collision avoidance manoeuvre will be the one with the lowest level of privilege, as it will be easier for it to move than the other one, which will be considered as an obstacle. The rules that are applied during the scenario (and a reminder of full names) are highlighted as shown in Table \ref{fig.rules}.

\begin{table}
\setlength{\tabcolsep}{1pt}
    \centering
    \begin{tabular}{|m{0.4\linewidth}|m{0.13\linewidth}|m{0.13\linewidth}|m{0.13\linewidth}|m{0.13\linewidth}|}
    \hline
         Rules :& 111 & 222 & 333& 444\\ \hline
         Finish OT&  & \cellcolor[gray]{.7} &  & \\ \hline
         Port OT& \cellcolor[gray]{.7} &  &  & \\ \hline
         Starboard OT&  &  &  & \\ \hline
         R to R&  &  &  & \\ \hline
    \end{tabular}

\small
    \begin{tabular}{|m{0.7\linewidth}|m{0.21\linewidth}|}
    \hline
    Finish overtaking the obstacle & Finish OT \\ \hline
    Overtaking the obstacle on the port side & Port OT\\    \hline
    Overtaking the obstacle on the starboard side & Starboard OT\\ \hline
    Red to Red rule to avoid collision & R to R\\    \hline
    \end{tabular}
 
    \caption{The rules grid is displayed alongside the main simulation to visualise which COLREGs'rules are applied. The upper numbers are the agent MMSI numbers.}
    \label{fig.rules}
\end{table}


The very first collision avoidance system consists of a potential based method and an associated controller. The collision zone is divided into four parts, and an adapted artificial potential field for each sector is applied. This potential field is updated every iteration depending on the new situation. This system is coupled with a path planning method. These two algorithms will be updated with the output of the research project according to module 3.

The AIS data ASV will move according to their AIS data located in .csv files, and the added ASV will have a final destination automatically calculated depending on their initial heading.

Multiple investigations can be considered using the simulator such as:
\begin{itemize}
    \item \textit{Testing acceptability}: different kind of situations can be run, which will each illustrate a situation where a collision is possible without any sort of intervention to correct it. The system will be acceptable if the potential collisions are avoided, and if the chosen new trajectory is realistic and validated by the AAA process. In Figure \ref{fig.rule14}, the initial path is represented by a purple dotted line. The red circle represents the distance to the closest point of approach (DCPA); in other words, it is the area which must not be overcome by any sort of obstacle. For the magenta circle, it represents the manoeuvring zone, where boats will have to start the avoiding collision manoeuvres also in respect of the COLREGs rules. The result is the trace of a new green line, which represents the corrected chosen path.
    
\begin{figure}
    \centering
\includegraphics[width=0.495\linewidth]{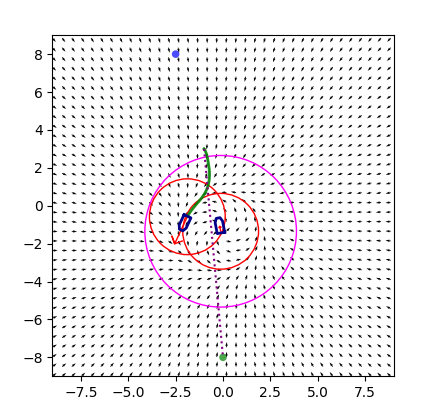} 
\includegraphics[width=0.495\linewidth]{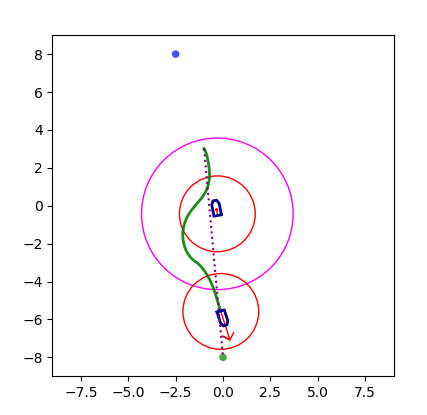} 
\caption{Illustration of Rule 14 - Head-on situation from the COLREGs implemented in the simulator.}
\label{fig.rule14}
\end{figure}

    \item \textit{Testing feasibility}, any new collision avoidance or path planning algorithm can be tested. As an example, the current chosen path planning algorithm could easily be replaced by another one, like the Bézier polynomials method or any AI-based algorithm, to study an other situation where obstacles would be registered. As a light simulator, it has been built to be suitable for long learning/training. 
\end{itemize}

\section{Conclusion}

The Convention on the International Regulations for Preventing Collisions at Sea, or COLREGs, primarily focus on rules for safe navigation at sea, assuming that all vessels share the common goal of avoiding collisions. However, it is important to consider the context in which these regulations operate, especially in relation to human navigators and their behaviour.
\begin{itemize}
    \item While COLREGs provide a framework for safe navigation, understanding the motivations and decision-making processes of human navigators can be crucial in ensuring compliance with these regulations. There may be cases where better understanding and modelling of human navigators can lead to improved outcomes, such as fewer collisions and near misses.
    \item In some situations, sailors may find themselves in mixed-motive scenarios where compliance with COLREGs conflicts with other goals or motivations, such as the desire to minimise time to destination or fuel consumption. Understanding these competing motives and how they influence human navigators' decisions can help design better training programmes, develop more effective autonomous systems.
    \item Understanding human motivations and concerns is also relevant in the context of autonomous navigation. Many stakeholders, including shipowners, captains, and crew members, may have reservations about adopting autonomous navigation systems. By considering and addressing these concerns, such as trust in technology, it may be possible to increase the acceptance and adoption of autonomous navigation solutions, which in turn could contribute to safer and more efficient maritime operations.
\end{itemize}
The objective is to get a deeper understanding of human behaviour and motives and to find out how it can contribute to safer and more effective maritime operations. It can also bridge the gap between regulatory frameworks such as COLREGs and the practical challenges of ensuring compliance and safety in real-world navigational scenarios.

This paper also proposed a combined approach to advance autonomous maritime navigation through the development of multi-constraint optimisation-based planners involving the formalisation of the human understanding of COLREG rules. The primary objective is to identify long-term COLREGs-compliant trajectories with a high navigation success rate for autonomous surface vessels (ASVs) while ensuring safe encounters with both manned and unmanned vessels within the region.

This research focuses on demonstrating the effectiveness of the proposed methods in single-ship encounters and benchmarking its performance against state-of-the-art techniques in multi-ship scenarios. The success rates obtained from extensive simulations serve as crucial indicators of the efficacy of the approach, but they are not the final aim.

\bibliography{aaai_23_bc}

\end{document}